\documentclass[preprint,12pt,3p,authoryear]{elsarticle}

\usepackage[title]{appendix}

\usepackage{natbib}

\usepackage{amssymb}
\usepackage{subfig}
\usepackage{color}
\usepackage{hyperref}
\usepackage{bm}
\usepackage{amsmath}
\usepackage{multirow}

\usepackage{setspace}

\begin{document}

\begin{frontmatter}

\title{\large{
Aggregating multiple types of complex data in stock market prediction:
\\A model-independent framework
}}
\author[label1,label2]{Huiwen Wang}
\author[label1,label3]{Shan Lu}
\author[label1,label2]{Jichang Zhao\corref{cor1}}

\address[label1]{School of Economics and Management, Beihang University}
\address[label2]{Beijing Advanced Innovation Center for Big Data and Brain Computing}
\address[label3]{Beijing Key Laboratory of Emergency Support Simulation Technologies for City Operations}
\cortext[cor1]{Corresponding author: jichang@buaa.edu.cn}

\begin{abstract}
The increasing richness in volume, and especially types of data in the financial domain provides unprecedented opportunities to understand the stock market more comprehensively and makes the price prediction more accurate than before. However, they also bring challenges to classic statistic approaches since those models might be constrained to a certain type of data. Aiming at aggregating differently sourced information and offering type-free capability to existing models, a framework for predicting stock market of scenarios with mixed data, including scalar data, compositional data (pie-like) and functional data (curve-like), is established. The presented framework is model-independent, as it serves like an interface to multiple types of data and can be combined with various prediction models. And it is proved to be effective through numerical simulations. Regarding to price prediction, we incorporate the trading volume (scalar data), intraday return series (functional data), and investors' emotions from social media (compositional data) through the framework to competently forecast whether the market goes up or down at opening in the next day. The strong explanatory power of the framework is further demonstrated. Specifically, it is found that the intraday returns impact the following opening prices differently between bearish market and bullish market. And it is not at the beginning of the bearish market but the subsequent period in which the investors' ``fear" comes to be indicative. The framework would help extend existing prediction models easily to scenarios with multiple types of data and shed light on a more systemic understanding of the stock market.
\end{abstract}


\end{frontmatter}
\newpage
\doublespacing

\section{Introduction}

Predicting stock prices have attracted significant research interests in both theories and applications. With the development of technology, the data related to stock market has been increasingly accumulated and diversified in either the sources or the types. For example, the direct sources originate in the financial system itself, such as the price information at various frequencies \citep{harris1986transaction,jain1988dependence,pan2017multiple}, the companies financial reports \citep{jones1970quarterly,zhou2015performance,zhou2017predicting}, and financial news \citep{geva2014empirical,li2014effect,hagenau2013automated}. The indirect sources are those outside the financial system, like the rise and fall of macro economic \citep{chen1986economic}, the reactions and reflections from investors' emotion revealed by social media \citep{zhou2017,sun2017predicting,ruan2018using,zhang2017improving,li2014effect}, search engine \citep{preis2013quantifying} or analyst's recommendations \citep{duan2013posterior}, etc. The richness of data sources has provided the chances to understand the stock market more comprehensively and make the price prediction more accurate than before. In the meantime, it brings challenges to the classic statistic analysis since they may not be suitable for dealing with these data at the same time. For example, the investors' emotion data are pie-like whose components are the proportions of different emotions, however, the intraday return series are curve-like. The former type of data is usually considered as compositional data and the latter one is functional data. As the two types of data belong to different spaces, it is not reasonable to directly combine them together and deploy statistic analysis.

The complex data analysis has been rapidly developed in the past decades. Two of the most popular types of complex data are compositional data and functional data. One observation of the compositional data is constituted by $D$ proportions that subject to a unit sum constraint.
Since 1896, compositional data has been the focus of research ~\citep{pearson1896mathematical,chayes1960correlation}, and has been applied in many research fields such as economics~\citep{longford2006stability}, ecology~\citep{aebischer1993compositional,bingham2007misclassified}, geochemistry~\citep{buccianti2006compositional,miesch1977log}, social science \citep{godichon2018clustering}, etc. Meanwhile, the studies on functional data analysis (FDA) has grown rapidly  \citep{ramsay1997functional,ferraty2006nonparametric,horvath2012inference,
fan2015functional} in the past decades. One observation of the functional data consists of a function (often a smooth curve, but not always). Functional linear model is among the most popular methods that have been widely used in the FDA~\citep{ramsay2007applied,horvath2012inference,cai2006prediction}. Many results have been published on the functional linear model, in which only functional predictor is presented \citep{hall2016truncated,comte2012adaptive,garcia2014goodness,escabias2004principal,
shang2015nonparametric,huang2016robust}. Nevertheless, these regression models are constrained to only single type of data and there are few efforts that consider the mixed-type of data when modelling \citep{wang2016generalized}. And in fact, aggregating data of various sources or formats will indeed enrich the views and resolutions of the explorations.

Stock prices prediction is one of such cases. While the daily price series are the most common data when conducting prediction \citep{hsu2011hybrid,jasemi2011modern,efendi2018new,baralis2017discovering,ye2016novel,chen2015hybrid}, other kinds of data are attracting people's attention as rapid growth of the generation and reservation of financial data. Firstly, public online emotion from social media has been used in predicting the stock market. Based on sixty thousands of microblogs from Sina Weibo, the largest online social media in China, \citet{wanyun2013investors} demonstrates that the public online emotion could only predict the trading volume rather than the prices. However, since they conduct the study using neither a huge amount of microblogs nor an effective classifier, their results may not be a generic conclusion. \citet{zhou2017} assigns five labels, including ``anger'', ``disgust'', ``joy'', ``sadness'', ``fear'', to over 3.5 million microblogs and shows that ``disgust'', ``joy'', ``sadness'', ``fear'' could be useful in predicting the Chinese stock market index. The daily emotions of the five types are naturally compositional data observations.
 Secondly, as the time series of intraday prices at various frequency became available, researchers have documented many intraday phenomena that related to stock returns, including the prices rise at the end of the day \citep{harris1986transaction,harris1989day}, significant weekday differences in intraday returns accrue during the first 45 minutes after the market opens \citep{harris1986transaction}, largest stock returns occur during the first (except on Monday) and the last trading hours, the lowest average return is earned in the fifth hour of the day \citep{jain1988dependence}. However, to our best knowledge, the exact knowledge on how the intraday returns would influence the future price is still to be discussed except one existing study that endorses autoregressive, random walk linear models, smooth transition, Markov switching, artificial neural network, non-parametric kernel regression and support vector machine models to predict the intraday returns \citep{matias2012forecasting}. Considering the inconsistent frequencies of the intraday return series with the commonly used daily return series, we argue that the intraday return curves can be used as observations of functional data, that is, one curve for one trading day. Thirdly, daily trading volume has been proved to be an important indicator in stock analysis as it is used to measure the relative worth of a market move \citep{foster1993variations,lillo2003econophysics}, and it usually belongs the common scalar data. Given the abundant information of stock market, however, how to integrate the multiple types of data into one prediction model is still unknown but of great importance for understanding and predicting the stock market.

To fill this vital gap, in this study, we propose a framework that incorporates the investors' emotions from social media (compositional data), intraday return series (functional data), and trading volume (scalar data) together to predict whether the market goes up or down at opening in the next day. As the goal of prediction is binary, it could be viewed as a classification problem. 
Specifically, by transforming the original data in terms of isometric logratio transformation, and functional principal component basis expansion, respectively, we can sufficiently obtain consistent numeric types of features from both compositional data and functional data. Since the transformation is independent to the prediction classifier, the framework serves as an interface between the data and the prediction model. We adopt logistic regression as a case of the classification models and present the corresponding estimation procedure. Note that other classification models could also be combined into the present approach, while logistic regression is particularly useful when the class is dichotomous and it does not need to assume data distributions on variables. More importantly, unlike the ``black-box'' approaches such as support vector machine, logistic regression could provide the predictors' coefficients, which is important for the model to give more insights on the relationships between the predictors and the response. Due to these benefits, we mainly consider logistic regression in this paper, but the framework can be combined with any prediction model. The estimation procedure of the framework is further proved to be consistent and effective through numerical simulations. 

In the real-world application, by dividing the sample period into three phases, the model exhibits a good prediction power when conducting out-of-sample predictions, especially in the first two phases. Besides, we find that both functional coefficients and numeric coefficients shed light on the different market status.
Most surprisingly, we find that in the bullish market (phase 1), the ``sadness" is more indicative than ``joy". And in the initial market crash (phase 2), the ``disgust" plays a dominant role in explaining the market. When the market became depressed (phase 3), the ``anger" and ``fear" begun to do their parts as well as other emotions.
Furthermore, our results show that it is not at the beginning of the bearish market but the subsequent period in which the investors' ``fear" comes to be indicative.

The rest of the paper is organized as follows. Section \ref{sec:data} introduces the data of the three types of predictors and the binary response that have motivated us to develop the framework with multiple types of data. In section \ref{sec:model}, we illustrate the transformation approaches to deal with compositional data and functional data, and then propose the model-independent framework. We also present how to estimate parameters of the logistic regression under the framework, which is considered as a case of the classification methods. Section \ref{sec:simulation} performs the simulation studies to prove that the proposed framework could yield effective estimation results. Section \ref{sec:result} presents the results of the prediction on Chinese stock market index, from the view of both explanation and prediction power. In section \ref{sec:conclusion}, we draw the conclusions as well as some of the limitations of this paper.

\section{Data}\label{sec:data}

\subsection{Sample period and binary response}\label{subsec:sample}
\begin{figure}[h!]
\centering
\includegraphics[width=1\linewidth]{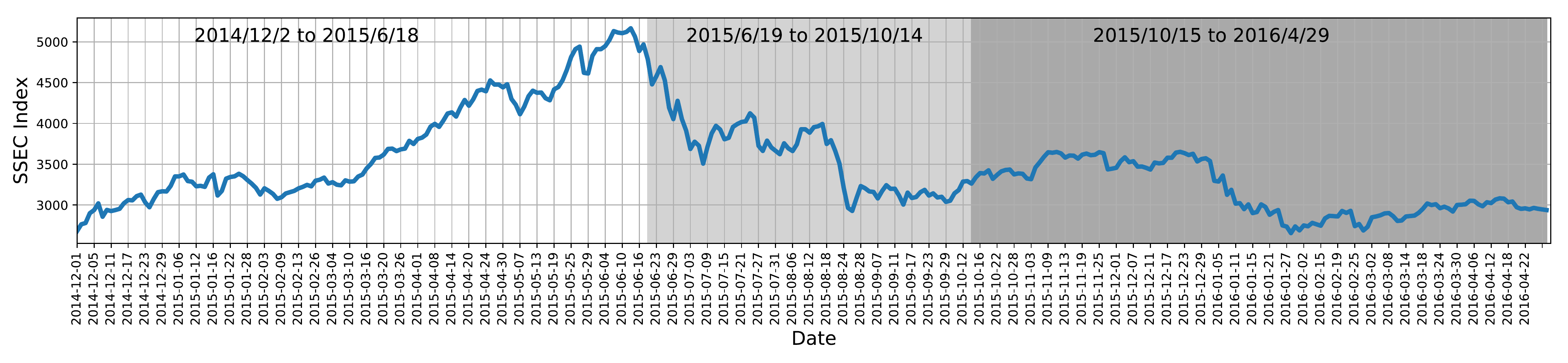}
\caption{{\bf Shanghai Stock Exchange Component index.} The sample period of this study starts from 2014/12/02 to 2016/4/29 (345 trading days in total), and completely covers the recent boom and bust of the Chinese stock market. The sample period is divided into three phases illustrated by different backgrounds.}\label{fig:market_index}
\end{figure}

In this study we consider the Chinese stock market, one of the largest markets in the world based on market capitalization. The sample period of this study is 2014/12/02 to 2016/4/29 (345 trading days in total), covering the recent boom and bust of Chinese stock market. As can be seen in Fig.~\ref{fig:market_index}, Shanghai Stock Exchange Composite index, which is one of the most important stock market indices in China, kept rising from the end of 2014 to the top of the past seven years at June 2015, and then went down sharply in the following months. From then on, the market kept vibrating at the low level that is close to the end of 2014.

Since the underlying market fundamentals vary a lot across the whole period, we cut the sample period into three phases illustrated by different backgrounds in Fig.~\ref{fig:market_index}. The first phase starts from 2014/12/2 to 2015/6/18, which had witnessed the enormous booming of the Chinese stock market. The second phase ranges from 2015/6/19 to 2015/10/14, starting with the popping of the stock market bubble and followed by severe turbulence though government had implemented a lot of bailout measures. The third phase defined in this paper starts from 2015/10/15 to 2016/4/29, when the market suffered from the major systemic aftershock and kept being depressed. Moreover, as will be discussed in section~\ref{subsec:com_data}, the pattern of the market emotions varies with the three phases. Therefore, the following analysis and modelling are applied to the three phases separately.

In this paper, the Shanghai Stock Exchange Composite (SSEC) index is employed as an indicator to represent the trend of Chinese stock market. The closing price of SSEC on day $i-1$ is denoted as $closing_{i-1}$ and the opening price of SSEC on day $i$ is denoted as $opening_{i}$. Then the daily open return of SSEC on day $i$ is defined as $y_i^\ast=(opening_{i}-closing_{i-1})/closing_{i-1}$. The reason is that this kind of percentage change is consistent with what investors see at the moment of the market opening on any trading information board \citep{lu2017herding}. Instead of the exact value of the open return, whether the open return is positive or negative is of the foremost interest in reality, because it could provide advice on the direction of trading. Therefore, using zero as the cut point, the $y_i^\ast$ is transformed into a binary variable $y_i$, i.e.

\begin{equation}
y_i=
\left\{
\begin{array}{rcl}
    1, &   &   {y_i^\ast>0}\\
    0, &    &{otherwise}.
\end{array}
\right.
\end{equation}
$y_i, i=1,...,n$ is used as the binary response in section \ref{sec:result}. 

\subsection{Functional predictor: intraday returns}\label{subsec:intraday}

There are 4 trading hours for continuous auction in one typical trading day in Chinese stock market, from 9:30am to 11:30am and 13:00pm to 15:00pm. To depict the intraday returns of SSEC, we use the last price of every five minutes in order to calculate the intraday percentage changes. Denote the price of SSEC at time $t$ on day $i$ as $p_{i,t}$, then the intraday return of SSEC at time $t$ on day $i$ is $r_{i,t}=(p_{i,t}-p_{i,t-1})/p_{i,t-1}$, as commonly defined in most financial studies. Starting from 9:35am to 11:30am and 13:00pm to 15:00pm, 49 points are included in one observation (one trading day). The series of intraday returns are treated as functional data because they could provide the consecutive information on trending of the index as a curve. The relative techniques to deal with functional data are introduced in section \ref{subsec:deal_with_function}.

As there are 345 trading days in our sample, 345 curves of the intraday returns are included in the study, one for each trading day, denoted as $x_i(t), i=1,...,n$. $x_i(t)$ is the functional predictor in section \ref{sec:result}.

\subsection{Scalar predictor: volume}\label{subsec:volume}

Volume is an important indicator in stock analysis as it is used to measure the relative worth of a market move \citep{foster1993variations,lillo2003econophysics}. Denote the volume on day $i$ in Shanghai Stock Exchange as $z_i, i=1,...,n$ and we use it as the scalar predictor in section \ref{sec:result}.

The data introduced in section \ref{subsec:sample}, \ref{subsec:intraday}, \ref{subsec:volume}, are downloaded from Thomson Reuters' Tick History.

\subsection{Compositional predictor: market emotion}\label{subsec:com_data}
\begin{figure}[h]
\centering
\includegraphics[width=1\linewidth]{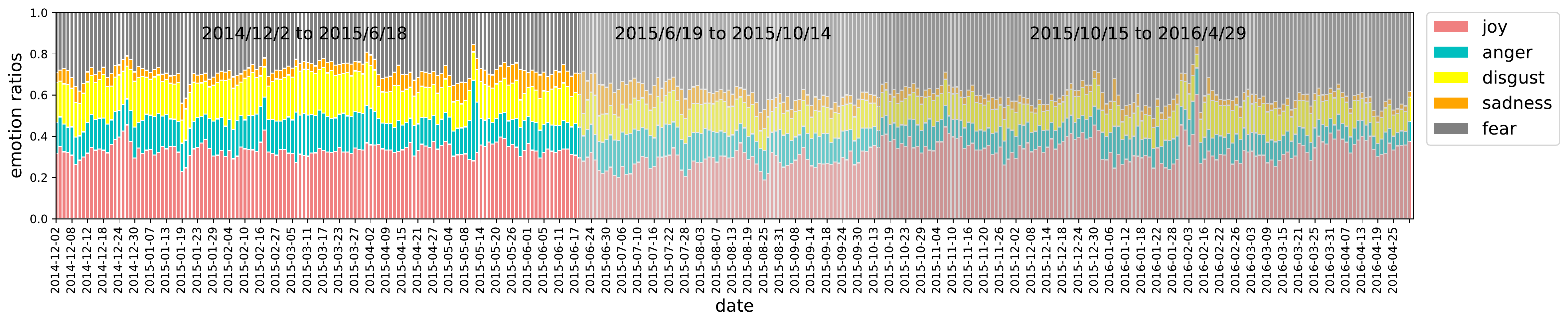}
\caption{{\bf Market emotions from Sina Weibo \citep{zhou2017}.} The sample period is divided into three phases illustrated by different backgrounds, as in Fig.~\ref{fig:market_index}.}\label{fig:emotion}
\end{figure}

Recent studies have found that the relationship between social media sentiments and stock returns is time-varying \citep{ho2017time} and some have successfully incorporated investors' emotions from social media into predicting stocks' prices \citep{zhou2017,sun2017predicting,ruan2018using,li2014effect}.
In this paper, we use the emotion measures from \citet{zhou2017}. Based on over 3.5 million emotionally labelled tweets from Sina Weibo as the corpus and a fast Naive Bayes classifier \citep{zhao2012moodlens}, they have arranged the daily stock-relevant tweets into five categories, namely ``anger", ``disgust", ``joy", ``sadness", and ``fear".
By scaling each kind of emotion by the sum of tweets in a day, we obtain the daily emotion ratios, respectively, to represent the investors' emotion ratios towards the market. The data are naturally compositional, each observation (one trading day) has five parts and the five parts add up to 1, denoted as $\bm{c_i}=[c_{i1}, c_{i2},c_{i3},c_{i4},c_{i5}], i=1,...,n$. Fig.~\ref{fig:emotion} illustrates the daily market emotion ratios over the sample period. The ``fear'' and ``joy'' had kept been the two dominant feelings among the five types of market emotion. Specifically, in the first phase, ``joy" is the largest part of the market emotions and ``fear'' is the second. In the second phase, however, the two kinds of emotions switch their positions, ``fear" became to be the largest. In the third phase, ``joy'' and ``fear'' alternatively dominates the market emotions. In addition, ``fear'' began to grow while ``disgust'' shrank a little in the bearish market (phase 2 and phase 3). The dominant roles of ``fear'' and ``joy'' in the compositions, however, does not mean that they are consequentially of the most indicative to the market trend. Section \ref{sec:result} will discuss this issue in details.

\section{Methodology}\label{sec:model}

In this section, we first introduce how to deal with compositional data and functional data using transformation techniques. Then, a model-independent framework for forecasting stock market is proposed based on these preliminaries. Finally, we use logistic regression model as a case of the framework and present the corresponding estimation procedure.

\subsection{Preliminaries}

\subsubsection{Compositional data}\label{subsec:deal_with_composi}
One observation of the compositional data is usually represented in the $D$-part simplex
\begin{equation}
\mathcal{S}^D=\{\bm{c_i}=[c_{i1},...,c_{iD}]\in \mathbb{R}^D | \sum_{j=1}^D c_{ij}=1,0<c_{ij}<1, \forall j\}.
\end{equation}
Since the natural constraints of compositional data, employing standard linear regression analysis for compositional data often leads to undesirable properties~\citep{aitchison1986statistical,pawlowsky2015modeling}.
A general solution is to remove the constraints first, and then apply classical statistics analysis on the transformed data to obtain the estimated coefficients. The last step is to transform the estimated coefficients back into the original simplex space. Under this circumstance, the key issue concerns about how to remove the constraints of compositional data by some transformation techniques before building the model. Many efforts have been devoted to investigate such kind of approaches. For instance, the additive logratio transformation \citep{aitchison1986statistical}, the centered logratio transformation \citep{aitchison1986statistical}, and the isometric logratio (${\rm ilr} $) transformation \citep{egozcue2003isometric}. Given the fact that ${\rm ilr} $ transformation is an isometry between $\mathcal{S}^D$ and $\mathbb{R}^{D-1}$, we use it to deal with compositional data in this paper.

For any $\bm{c_i}= [c_{i1}, c_{i2},\cdots, c_{iD}]\in \mathcal{S}^D$,  the ${\rm ilr} $ transformation maps $\bm{c_i}$ to $\bm{{c}_i^{\ast}}={\rm ilr} (\bm{c_i})=({c}_{i1}^{\ast}, {c}_{i2}^{\ast},\cdots, {c}_{i,D-1}^{\ast})\in \mathbb{R}^{D-1}$ by

\begin{equation}\label{eq:compo_transformation}
\bm{{c}_{i}^{\ast}} =  {\Big (}\ln \frac{{c}_{i1}}{\sqrt[D]{\prod_{j=1}^D {c}_{ij}}},\ln \frac{{c}_{i2}}{\sqrt[D]{\prod_{j=1}^D {c}_{ij}}},\cdots,\ln \frac{{c}_{iD}}{\sqrt[D]{\prod_{j=1}^D {c}_{ij}}} {\Big )}\bm{\Psi}',
\end{equation}
where $\bm{\Psi}$ is a $(D-1,D)$ matrix
\begin{equation}\label{eq:ilr_basis}
\Psi_{ij}=
\left\{
\begin{array}{ccl}
    +\sqrt{\frac{1}{(D-i)(D-i+1)}}, &   &   {j\leq D-i,}\\
    -\sqrt{\frac{D-i}{D-i+1}},& &   {j=D-i+1,}\\
    0, &    &{otherwise,}
\end{array}
\right.
\end{equation}
as \citet{egozcue2003isometric} proposed.

\subsubsection{Functional data}\label{subsec:deal_with_function}

In recent years, FDA has been rapidly developed \citep{ramsay1997functional,ramsay2007applied,ferraty2006nonparametric,horvath2012inference}.
When it comes to functional linear regression, for the case of a continuous scalar response variable and a functional predictor for the individual of interest  \citep{ramsay1997functional}, both non-data-driven basis (e.g. B-spline) and data-driven basis (e.g. functional principal component) are commonly used. Particularly, \citet{hall2007methodology} considers the least square estimator for functional linear regression model based on functional principal components and obtains the optimal convergence rate of the slope function. \citet{meng2016comparison} pointed out that functional principal component basis could be the first preference, especially as we have none prior knowledge on the functional data types. Thus, we consider dealing with functional variables based on functional principal components basis expansion.

Define the covariance function of $X(t)$ by $K(s,t)={\rm cov}[X(s),X(t)]=E[X(s)X(t)],$ then by Mercer's Theorem we can obtain the spectral decomposition
\begin{equation}\label{eq:spectral}
K(s,t)=\sum_{j=1}^{\infty}\theta_{j}\phi_{j}(s)\phi_{j}(t),
\end{equation}
where $\theta_{1}\geq\theta_{2}\geq\cdots\geq0$ are eigenvalues of the operator associated with $K(s,t)$, and $\{\phi_{j}\}$ are the corresponding eigenfunctions. According to the Karhunen-Loeve representation, we have $X(t)$ in the space of $\{\phi_{j}\}_{j=1}^{\infty}$ as
\begin{equation}\label{eq:inf_expansion}
X(t) = \sum_{j=1}^{\infty}a_j\phi_j(t).
\end{equation}

Assume that we have independently identically distributed (i.i.d.) observations $\bm{X}(t)=(x_{1}(t),\cdots,x_{n}(t))$, where $n$ denotes the sample size. Recall that $\bm{X}(t)$ is assumed to be zero-mean, the empirical covariance function is
\begin{equation}\label{eq:covariance}
\hat{K}(s,t)=\frac{1}{n}\sum_{i=1}^nx_{i}(s)x_{i}(t),
\end{equation}
which can be used to estimate $K(s,t)$. Same with Equation (\ref{eq:spectral}) we can obtain that
\begin{equation}\label{eq:sample_pca}
\hat{K}(s,t)=\sum_{j=1}^{\infty}\hat{\theta}_{j}\hat{\phi}_{j}(s)\hat{\phi}_{j}(t),
\end{equation}
where $\hat{\theta}_{1}\geq \hat{\theta}_{2}\geq \cdots\geq 0.$ Usually $\int \hat{\phi_j}(t)\hat{\phi}_j (t)dt>0$ is assumed to get rid of uncertainty of signs.
Considering $\{\hat{\phi}\}$ is a basis of the spanned space by $\bm{X}(t)$, we see that at most $n$ eigenvalues are strictly positive. Then we obtain
\begin{equation}\label{eq:funx_expasion}
 \bm{X}(t) \approx \sum_{j=1}^{M} \bm{a}_j \phi_j(t),
\end{equation}
where $M$ is the number of the basis functions, and $M$ is usually determined by
\begin{equation}\label{Eq:m_max}
M=\underset{1\leq m \leq n}{\rm{arg}\min}\{\sum_{i=1}^{m}\hat{\theta}_i/(\sum_{i=1}^n\hat{\theta}_i)\geq \lambda\},
\end{equation}
where $\lambda$ is usually set to be 85\% \citep{wang2016generalized}.
From Equation (\ref{eq:funx_expasion}), the functional subspace $\mathcal{L}^2$ is spanned by the set of orthonormal basis $\{\hat{\phi}\}$ to M-dimensional real space and accordingly $(\bm{a}_1, \bm{a}_2,..., \bm{a}_M)$ can be used to represent $\bm{X}(t)$.

\subsection{The framework of aggregating multiple types of complex data}
\begin{figure}[h!]
\centering
\includegraphics[width=1\linewidth]{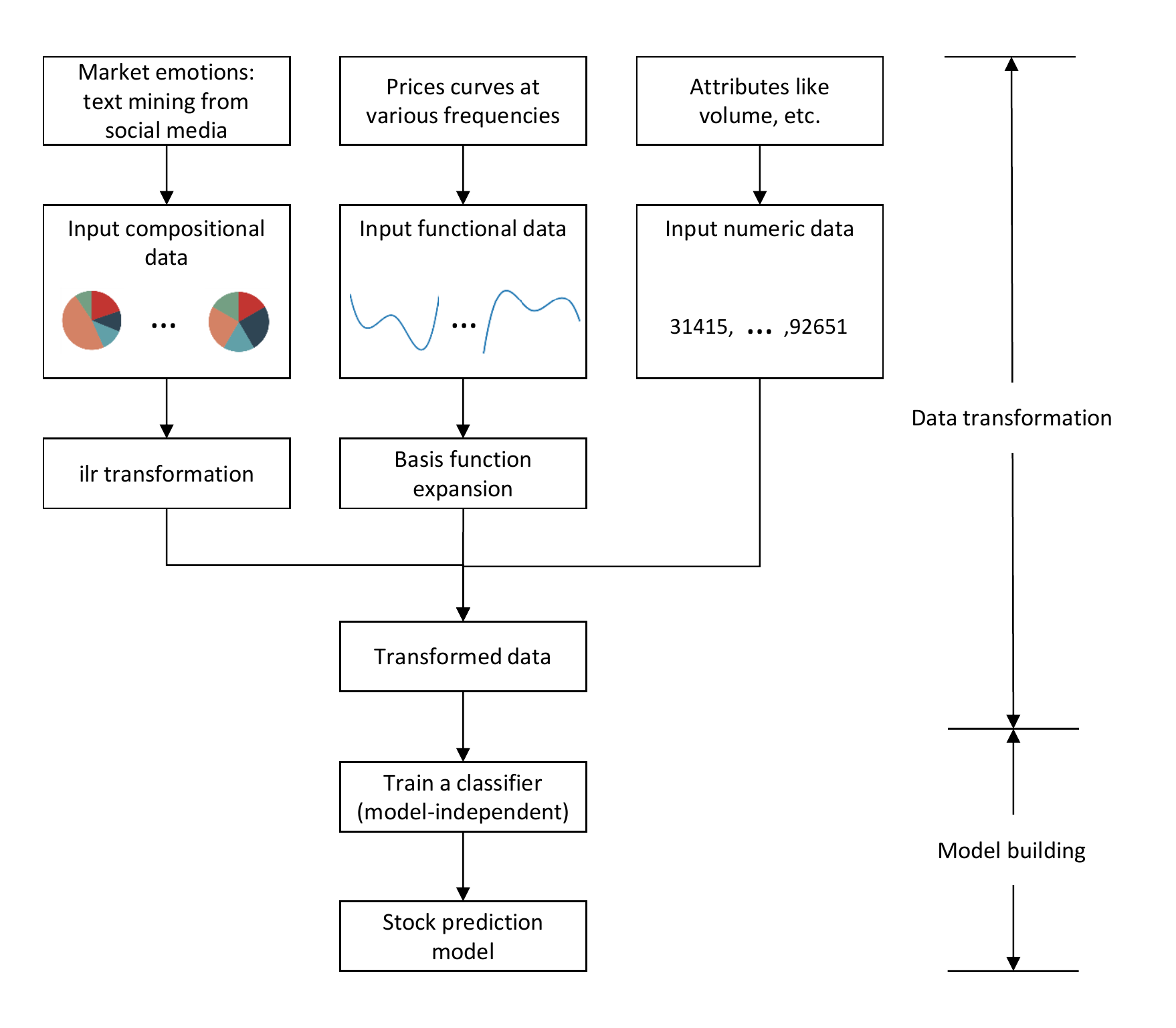}
\caption{{\bf The framework for stock prediction using mixed types of complex data.}}\label{fig:framework}
\end{figure}

Based on the above, we propose a framework for stock prediction using mixed types of complex data, as shown by Fig.~\ref{fig:framework}. The massive emotional information embedded in social media could be converted into proportions of different kinds of feelings and integrated into the framework as compositional data. The intraday returns could be viewed as curves and delivered to the framework as functional data. Other attributes like daily trading volumes are scalar data. The isometric logratio transformation, functional principal component basis expansion are used to reconstruct the original complex data to provide equivalent numerical transformed data for further statistical analysis. Then, the logistic regression classifier is trained with the transformed data, and the prediction model is thus built. Note that based on the transformed data, any model, either the regression ones or machine learning ones, can be trained to perform the prediction. That's to say, our framework is model-independent and offers an interface of aggregating multiple types of complex data. The prediction model could be used to predict the newly arrived data and give its opinion of whether the market will go up or down.

It is worth noting that the transformation procedures is essentially space transformation, that is, from simplex to real space for compositional data, from Hilbert space to real space for functional data. The isometric logratio transformation fully retains the information of the original compositional data, while the basis function expansion based on functional principal components basis expansion absorbs the most informative elements from the functional data. Therefore, the transformed data well represent the original data on the whole.

Furthermore, suppose that a data set has $N$ samples, each of the functional data sample has $p$ observation points, and each of the compositional data sample has $D$ parts. Then the time complexity of basis function expansion based on functional principal analysis is $\mathcal{O}(Np^2)$ as it is necessary to compute the covariance function in Equation (\ref{eq:covariance})~\citep{meng2016comparison}, while the time complexity of $ilr$ transformation is $\mathcal{O}(ND^2)$ as it transforms the compositional data sample one-by-one using a matrix of $D\times(D-1)$ dimensions as shown in Equation (\ref{eq:ilr_basis}). Considering that usually $p<100$ and $D<10$ in real-world applications, in the big data era, the cost for these transformations is rather small. From this perspective, the framework offers a simple but effective solution to deal with aggregated, big, complex data in the finance domain.

More importantly, the framework is model-independent. It offers an interface of aggregating multiple types of complex data. Based on the transformed data, any kind of prediction models can be accordingly trained. Classification methods except logistic regression are good alternatives in the model building stage, but we choose logistic regression for the rest of discussion. The reason is that logistic regression not only is particularly useful when the class is dichotomous but also provides the predictors' coefficients, which enables us to collect the intuitions buried in the model.

\subsection{Logistic regression and estimation for multiple types of complex data}\label{subsec:logistic}

In this section, we consider a logistic regression model with three kinds of predictors, including scalar data, functional data and compositional data. It is worth noticing that for classifiers like support vector machine, the compositional and functional data are feed into the model in their transformed forms by Equation (\ref{eq:compo_transformation}) and (\ref{eq:funx_expasion}). However, as logistic regression model involves the coefficients estimation, the inner products of coefficients and variables are required. Therefore, we present the details in the rest of the section.

For the sake of convenience, assume all the variables are zero-mean. Using the inner product expressions for compositional variable and functional variable, our model is

\begin{equation} \label{mdl}
Y = h(\gamma Z +  \langle \bm{\alpha}, \bm{C}\rangle_{\mathcal{S}}+\int X(t)\beta(t)dt + \epsilon),
\end{equation}
where $Z\in \mathbb{R}$ is a scalar variable, $\bm{C}\in \mathcal{S}^D$ is a compositional variable of $D$ parts, $X(t)\in \mathcal{L}^2$ is a zero mean, second-order stochastic process, $Y\in\{0,1\}$ is the binary response. $\langle \cdot , \cdot \rangle_{\mathcal{S}}$ is the corresponding inner product operator for the compositional variable. $\gamma \in \mathbb{R}, \bm{\alpha} \in \mathcal{S}^D$ and $\beta(t) \in \mathcal{L}^2$ are coefficients to estimate and $\epsilon$ is a random error term. As logistic regression takes into account features' joint effects and levels a good linear combination of features as the decision boundary, we consider $h(u)=\frac{\exp(u)}{1+\exp(u)}$ as the link function in this paper.

Note that, the dimension of $\gamma$ is fixed once $Z$ is given, while $\bm{\alpha}$ and $\beta(t)$ may be varying under different transformation methods. In the following subsections, we will deal with the compositional and functional data one by one, trying to transform them into scalar data so as to conduct regular statistics techniques. At the end of the section, we give the maximum likelihood estimation procedure using the transformed data. Hereafter, denote sample $i$ as $(y_i, z_i, \bm{c_i}, x_i(t)), i=1,2,...,n$.

According to section \ref{subsec:deal_with_composi}, by the ${\rm ilr} $ transformation, the inner product of $\bm{\alpha},\bm{c_i}\in\mathcal{S}^D$ is converted into
\begin{equation}
\langle \bm{\alpha}, \bm{c_i}\rangle_{\mathcal{S}} = \sum_{j=1}^{D-1}\alpha_{j}^{\ast}c_{ij}^\ast,
\end{equation}
where $\bm{\alpha}=[\alpha_1,\alpha_2,...,\alpha_D]\in\mathcal{S}^D$, $\bm{\alpha^{\ast}}={\rm ilr}(\bm{\alpha}) = (\alpha_1^{\ast},\alpha_2^{\ast},...,\alpha_{D-1}^{\ast})\in\mathbb{R}^{D-1}$.
Denote $\mathcal{C(\cdot)}$ as the closure for a composition to rescaling of the initial vector so that the sum of its components is 1, i.e. $\mathcal{C}(\bm{u})=[\frac{u_1}{\sum_{i=1}^D u_i},\frac{u_2}{\sum_{i=1}^D u_i},..., \frac{u_D}{\sum_{i=1}^D u_i} ]$. $\bm{\alpha^{\ast}} \in \mathbb{R}^{D-1}$ could be transformed back to $\bm{\alpha} \in\mathcal{S}^{D}$ by the ${\rm ilr}^{-1}$ transformation
\begin{equation}\label{eq:com_reverse}
\bm{\alpha} = {\rm ilr}^{-1}( \bm{\alpha^{\ast}} )=\mathcal{C}(\exp(\bm{\alpha^{\ast}} \bm{\Psi} )).
\end{equation}

According to section \ref{subsec:deal_with_function}, we have $X(t)$ and $\beta(t)$ in the space of $\{\phi_{j}\}_{j=1}^{\infty}$, i.e. $\beta(t)= \sum_{j=1}^{\infty}b_j\phi_j(t)$, similar to Equation (\ref{eq:inf_expansion}). Then, considering the orthogonal of the ${\phi_j(t)}$, the third term in Model (\ref{mdl}) can be rewritten as
\begin{equation}
\int X(t)\beta(t)dt = \sum_{j=1}^{\infty} a_j b_j.
\end{equation}

Again, we assume that $\bm{X}(t)=(x_{1}(t),\cdots,x_{n}(t))$ are independently identically distributed (i.i.d.) observations with zero-mean. By incorporating Equation (\ref{eq:sample_pca}), the group of expansion basis are obtained, and similar to Equation (\ref{eq:funx_expasion}), we have

\begin{equation}\label{eq:fun_reverse}
\beta(t)= \sum_{j=1}^{M}b_j\phi_j(t),
\end{equation}
where $M$ is the number of the basis functions, and is determined by Equation (\ref{Eq:m_max}).

Thus,
\begin{equation}
\int \bm{X}(t)\beta(t)dt \approx \sum_{j=1}^{M} \bm{a}_j b_j.
\end{equation}

Thus, Model (\ref{mdl}) can be rewritten as
\begin{equation}
Y \approx h( \gamma Z + \sum_{j=1}^{D-1} \bm{\alpha}_{j}^{\ast} \bm{c}_{j}^{\ast} + \sum_{j=1}^{M} \bm{a}_j b_{j}+ \epsilon),
\end{equation}
in which parameters to be estimated contain $\gamma, \bm{\alpha}^{\ast}, \bm{b}$.

For any sample $i$, let $\bm{b}= (b_{1}, \cdots, b_{M})',  \bm{a_i}=(a_{i1}, \cdots, a_{iM})' $,
$y_i \approx  h(\gamma z_i + \bm{c}_i^{\ast '} \bm{\alpha}^{\ast} +\bm{a}_i'  \bm{b} +\epsilon_i).$
The link function is set to be $h(u)=\frac{\exp(u)}{1+\exp(u)}$. For any $u\in \mathbb{R}$. The expectation of $Y$ is
\begin{equation}\label{eq:pi}
\pi_i=P(Y=1|z_i, \bm{c_i}, x_i(t))\\
=\frac{\exp(\gamma z_i + \bm{c}_i^{\ast '} \bm{\alpha}^{\ast} + \bm{a}_i' \bm{b} +\epsilon_i)} {1+\exp(\gamma z_i + \bm{c}_i^{\ast '} \bm{\alpha}^{\ast} + \bm{a}_i'  \bm{b} +\epsilon_i)}, i=1,...,n.
\end{equation}
The likelihood of is $L= \prod_{i=1}^n\pi_i^{y_i}(1-\pi_i)^{1-y_i}$ and we can obtain the estimators by maximum the log likelihood
\begin{equation}
(\hat{\gamma}, \bm{\hat{\alpha}}^{\ast}, \bm{\hat{b})} =\underset{\gamma,\bm{\hat{\alpha}}^{\ast}, \bm{\hat{b}}}{\arg \max} \{ \sum_{i=1}^n y_i(\gamma z_i + \bm{c}_i^{\ast '} \bm{\alpha}^{\ast} + \bm{a}_i' \bm{b})-\sum_{i=1}^n ln(1+\exp(\gamma z_i + \bm{c}_i^{\ast '} \bm{\alpha}^{\ast} + \bm{a}_i' \bm{b})) \}.
\end{equation}

The estimated $\bm{\hat{\alpha}}^{\ast}, \bm{\hat{b}}$ would then be inversely transformed back to their original space denoted as $\bm{\hat{\alpha}}$ and $\hat{\beta}(t)$ using Equations (\ref{eq:com_reverse}) and (\ref{eq:fun_reverse}).

\section{Effectiveness of the framework}\label{sec:simulation}
In this section, we perform simulation studies to verify the effectiveness of the proposed framework with finite sample size. Although the framework serves as an interface between multiple types of complex data and stock prediction, it is important to assess the effectiveness of the parameters estimated by the framework because the parameters could equip the framework with explanatory power.

Given that the data transformation is independent to the prediction model in the framework, the logistic regression is considered as a case of the classification methods to evaluate the robustness and usefulness of the framework, as discussed in section \ref{subsec:logistic}. The details of the simulation are described as follows.

There are three types of predictors on each individual of interest, i.e., scalar data predictor, compositional data predictor, and functional data predictor. For simplicity, all the predictors are scaled by their centers in order to be zero-mean. The data are generated from the following model

\begin{equation}
y_i = h(\gamma z_i +  \langle \bm{\alpha}, \bm{c}_i\rangle_{\mathcal{S}}+\int \beta(t)x_i(t)dt + \sigma e_i),
\end{equation}
where  $h(u)=\frac{\exp(u)}{1+\exp(u)}$ is the link function, $y_i$ is the 0-1 response, $z_i$ is the scalar predictor, $\bm{c_i}$ is the compositional predictor, $x_i(t)$ is the functional predictor, $e_i$ is the noise.

In the simulation, we first generate the predictors.
$z_i\in \mathbb{R}$ is generated from normal distribution with mean equal to 0 and standard deviation equal to 1. $\bm{c_i}$ is of compositional data with three parts and each part is uniformly distributed. $e_i$ is normally distributed. $\sigma$ controls the ratio of signal to noise and here we set $\sigma = $ 0.2, 0.4, and 0.6. Besides, the functional data $x_i(t)$ and its functional coefficients $\beta(t)$ are generated on the $T=100$ equally spaced grids on $[0,1]$ as \citep{hall2007methodology}
\begin{eqnarray*}
&&\beta(t) = \sum_{j=1}^{50} \beta_{j}\phi_j(t),\\
&&\beta_{1}=0.3, \beta_{j}=4(-1)^{j+1}j^{-2},j\geq 2,\\
&&\phi_j(t)=\sqrt{2}\cos(j\pi t),\\
&&x_i(t) = \sum_{j=1}^{50}\gamma_jZ_j\phi_j(t),\\
&&\gamma_j=(-1)^{j+1}j^{-a/2},a=1.1,Z_j\sim U[-\sqrt{3},\sqrt{3}].
\end{eqnarray*}

Without loss of generality, let $\gamma=1, \bm{\alpha}=[\alpha_1,\alpha_2,\alpha_3]=[0.3,0.5,0.2]\in \mathcal{S}^3$. Then the probabilities are calculated by
\begin{equation}
\pi_i=\frac{\exp(\gamma z_i +  \langle \bm{\alpha}, \bm{c}_i\rangle_{\mathcal{S}}+\int \beta(t)x_i(t)dt + \sigma e_i)}{1+\exp(\gamma z_i +  \langle \bm{\alpha}, \bm{c}_i\rangle_{\mathcal{S}}+\int \beta(t)x_i(t)dt + \sigma e_i)}, i=1,...,n.
\end{equation}
And we finally obtain $n$ values of the response $y_i$ by simulating observations of a Bernouilli distribution with probabilities $\pi_i$.

After generating the simulated data, we use the proposed estimated procedure to obtain the estimated value of $\hat{\gamma}, \bm{\hat{\alpha}}, \hat{\beta}(t)$. Here we set $\lambda=85\%$ in Equation (\ref{Eq:m_max}) to determine the number of basis. The ``generate data-estimate coefficients" procedure is repeat 200 times for every sample size setting, i.e. $n=100,200,500,1000,2000,5000,10000$.

To measure the performances of the estimation procedure, we introduce mean of integrated square error (MISE) and correlation of true $\beta(t)$ and $\hat{\beta}(t)$ as

\begin{eqnarray*}
&& MISE(\hat{\beta}) = \int_0^1 (\hat{\beta}(t)-\beta(t))^2dt,\\
&& cor(\hat{\beta}(t),\beta(t)) = \frac{\sum _{i=1}^{T}(\beta(t)_{i}-{\overline {\beta}}(t))(\hat{\beta}(t)_{i}-\hat{{\overline {\beta}}}(t))}{\sqrt{(\sum_{i=1}^{T}(\beta(t)_{i}-{\overline {\beta}}(t))^2} \sqrt{(\sum_{i=1}^{T}(\hat{\beta}(t)_{i}-\hat{{\overline{\beta}}}(t)))^2}}\\
\end{eqnarray*}
where $T=100$ is the number of equally spaced grids on $[0,1]$. The results of the averaged $MISE(\hat{\beta}(t))$ and $cor(\hat{\beta}(t),\beta(t)) $ over the 200 times of simulation are shown in Table~\ref{tab:cor} and Table~\ref{tab:mise}.

\begin{table}[h]
\centering
\caption{The correlation of estimated functional coefficients $\hat{\beta}(t)$ and true functional coefficients $\beta(t)$ and its standard deviation (in parentheses).}\label{tab:cor}
\begin{tabular}{rccccccc}
  \hline
 sample size  & 100 & 200 & 500 & 1000 & 2000 & 5000 & 10000 \\
  \hline
 $\sigma=0.2$ & $\underset{(0.049)}{0.943}$ & $\underset{(0.03)}{0.958}$ & $\underset{(0.019)}{0.962}$ & $\underset{(0.012)}{0.964}$ & $\underset{(0.009)}{0.967}$ & $\underset{(0.006)}{0.967}$ & $\underset{(0.004)}{0.967}$ \\
 \hline
$\sigma=0.4$& $\underset{(0.05)}{0.94}$ & $\underset{(0.03)}{0.952}$ & $\underset{(0.017)}{0.964}$ & $\underset{(0.012)}{0.965}$ & $\underset{(0.008)}{0.966}$ & $\underset{(0.005)}{0.967}$ & $\underset{(0.004)}{0.967}$ \\
\hline
 $\sigma=0.6$ & $\underset{(0.048)}{0.937}$ & $\underset{(0.032)}{0.951}$ & $\underset{(0.018)}{0.962}$ & $\underset{(0.013)}{0.966}$ & $\underset{(0.009)}{0.965}$ & $\underset{(0.005)}{0.966}$ & $\underset{(0.004)}{0.967}$ \\
   \hline
\end{tabular}
\end{table}

\begin{table}[ht]
\centering
\caption{ $MISE(\hat{\beta})$: the MISE of the functional coefficients and its standard deviations (in parentheses).}\label{tab:mise}
\begin{tabular}{rccccccc}
  \hline
sample size & 100 & 200 & 500 & 1000 & 2000 & 5000 & 10000 \\
  \hline
 $\sigma=0.2$ & $\underset{(0.206)}{0.263}$ & $\underset{(0.109)}{0.16}$ & $\underset{(0.058)}{0.122}$ & $\underset{(0.035)}{0.109}$ & $\underset{(0.026)}{0.099}$ & $\underset{(0.016)}{0.098}$ & $\underset{(0.012)}{0.097}$ \\
 \hline
 $\sigma=0.4$ & $\underset{(0.193)}{0.263}$ & $\underset{(0.097)}{0.172}$ & $\underset{(0.052)}{0.117}$ & $\underset{(0.036)}{0.108}$ & $\underset{(0.026)}{0.104}$ & $\underset{(0.017)}{0.1}$ & $\underset{(0.011)}{0.101}$ \\
 \hline
 $\sigma=0.6$ & $\underset{(0.486)}{0.287}$ & $\underset{(0.111)}{0.173}$ & $\underset{(0.053)}{0.126}$ & $\underset{(0.039)}{0.112}$ & $\underset{(0.028)}{0.112}$ & $\underset{(0.017)}{0.108}$ & $\underset{(0.012)}{0.106}$ \\
   \hline
\end{tabular}
\end{table}

\begin{table}[h]
\centering
\caption{The bias of scalar predictor's coefficient $\hat{\gamma}$ and its standard deviations (in parentheses).}\label{tab:gamma}
\begin{tabular}{rccccccc}
  \hline
 sample size & 100 & 200 & 500 & 1000 & 2000 & 5000 & 10000 \\
  \hline
$\sigma=0.2$ & $\underset{(0.39)}{0.083}$ & $\underset{(0.271)}{0.041}$ & $\underset{(0.154)}{-0.005}$ & $\underset{(0.113)}{-0.024}$ & $\underset{(0.069)}{-0.036}$ & $\underset{(0.048)}{-0.041}$ & $\underset{(0.029)}{-0.035}$ \\
\hline
 $\sigma=0.4$ & $\underset{(0.391)}{0.074}$ & $\underset{(0.263)}{-0.016}$ & $\underset{(0.149)}{-0.036}$ & $\underset{(0.103)}{-0.031}$ & $\underset{(0.075)}{-0.043}$ & $\underset{(0.046)}{-0.047}$ & $\underset{(0.035)}{-0.057}$ \\
 \hline
$\sigma=0.6$& $\underset{(0.413)}{0.004}$ & $\underset{(0.265)}{-0.005}$ & $\underset{(0.153)}{-0.058}$ & $\underset{(0.102)}{-0.072}$ & $\underset{(0.066)}{-0.076}$ & $\underset{(0.048)}{-0.085}$ & $\underset{(0.031)}{-0.085}$ \\
   \hline
\end{tabular}
\end{table}

\begin{table}[h]
\centering
\caption{The bias of compositional coefficients $\bm{\hat{\alpha}}$ and its standard deviations (in parentheses).}\label{tab:alpha}
\begin{tabular}{rcccccccc}
  \hline
 sample size &  & 100 & 200 & 500 & 1000 & 2000 & 5000 & 10000 \\
  \hline
 $\sigma=0.2$ & $\hat{\alpha_1}$ & $\underset{(0.1)}{0.001}$ & $\underset{(0.056)}{0}$ & $\underset{(0.034)}{-0.001}$ & $\underset{(0.024)}{0.003}$ & $\underset{(0.017)}{0.001}$ & $\underset{(0.011)}{0.002}$ & $\underset{(0.008)}{0.001}$ \\
  & $\hat{\alpha_2}$ & $\underset{(0.105)}{0}$ & $\underset{(0.073)}{-0.002}$ & $\underset{(0.043)}{0.002}$ & $\underset{(0.028)}{-0.006}$ & $\underset{(0.021)}{-0.003}$ & $\underset{(0.013)}{-0.005}$ & $\underset{(0.009)}{-0.005}$ \\
  & $\hat{\alpha_3}$  & $\underset{(0.072)}{-0.001}$ & $\underset{(0.044)}{0.002}$ & $\underset{(0.028)}{-0.001}$ & $\underset{(0.019)}{0.003}$ & $\underset{(0.014)}{0.002}$ & $\underset{(0.008)}{0.003}$ & $\underset{(0.006)}{0.004}$ \\
  \hline
 $\sigma=0.4$ &$\hat{\alpha_1}$  & $\underset{(0.091)}{-0.003}$ & $\underset{(0.06)}{-0.006}$ & $\underset{(0.034)}{0.003}$ & $\underset{(0.022)}{0.004}$ & $\underset{(0.017)}{0.001}$ & $\underset{(0.011)}{0.002}$ & $\underset{(0.007)}{0.002}$ \\
  & $\hat{\alpha_2}$  & $\underset{(0.113)}{0.011}$ & $\underset{(0.074)}{0.007}$ & $\underset{(0.044)}{-0.004}$ & $\underset{(0.03)}{-0.01}$ & $\underset{(0.022)}{-0.006}$ & $\underset{(0.012)}{-0.008}$ & $\underset{(0.009)}{-0.009}$ \\
  & $\hat{\alpha_3}$  & $\underset{(0.07)}{-0.008}$ & $\underset{(0.047)}{-0.001}$ & $\underset{(0.031)}{0.001}$ & $\underset{(0.021)}{0.006}$ & $\underset{(0.015)}{0.005}$ & $\underset{(0.009)}{0.006}$ & $\underset{(0.006)}{0.006}$ \\
  \hline
 $\sigma=0.6$ & $\hat{\alpha_1}$  & $\underset{(0.1)}{0.013}$ & $\underset{(0.055)}{0.005}$ & $\underset{(0.035)}{0.006}$ & $\underset{(0.026)}{0.003}$ & $\underset{(0.016)}{0.003}$ & $\underset{(0.011)}{0.004}$ & $\underset{(0.007)}{0.004}$ \\
  & $\hat{\alpha_2}$  & $\underset{(0.111)}{-0.014}$ & $\underset{(0.069)}{-0.012}$ & $\underset{(0.039)}{-0.017}$ & $\underset{(0.031)}{-0.013}$ & $\underset{(0.021)}{-0.012}$ & $\underset{(0.013)}{-0.014}$ & $\underset{(0.009)}{-0.013}$ \\
  & $\hat{\alpha_3}$  & $\underset{(0.072)}{0.001}$ & $\underset{(0.051)}{0.007}$ & $\underset{(0.028)}{0.011}$ & $\underset{(0.019)}{0.01}$ & $\underset{(0.013)}{0.009}$ & $\underset{(0.009)}{0.01}$ & $\underset{(0.006)}{0.009}$ \\
   \hline
\end{tabular}
\end{table}
%
%
From Table \ref{tab:cor}, we can see that the estimated parameters are indeed highly correlated with the true parameter. The MISE also shows the unbiased attribute of the estimation procedure in an aggregated level. Both the standard deviations of the two decrease when the sample size becomes larger. As for $\hat{\gamma}, \bm{\hat{\alpha}}$, their averaged bias and standard deviations are shown in Table~\ref{tab:gamma} and Table~\ref{tab:alpha}, which both exhibit only tiny bias. Looking at the estimation variance of the bias, we can see it does decrease when we increase the sample size.
In conclusion, the simulation study above has provided unbiased and consistent estimation results, which advertises the effectiveness of the proposed framework.

\section{Empirical studies on Chinese stock market}\label{sec:result}
Considering the low cost  and evaluated consistency of the framework in aggregating complex data, it can be used in realistic applications like stock prediction. As the Chinese stock market's fundamentals vary throughout the sample period, we cut the whole period into three phases, as shown in Fig.~\ref{fig:market_index}.
We use the volume, emotion, and intraday returns of day $i$ as the predictors, while the states of the open return (if the return is positive, then $y_i=1$, otherwise $y_i=0$) in day $i+1$ as the response.
In this case, the framework serves as a prediction method for the open return of the index. Here, when transforming the functional data, $\lambda$ is set to be 99\% in Equation (\ref{Eq:m_max}) to determine the number of basis so as to retain the majority of information that the intraday return series could offer. Furthermore, note that unlike machine learning methods like support vector machine, the advantage of logistic regression is that it could access the coefficients of the variables instead of focusing on the prediction power alone. As both the interpretation and the predicting power of the proposed framework are of interest, we present them one by one in the following sub-sections.

\subsection{Predication}

We perform the 5-fold cross validation for the three phases separately. The cross validation is a widely used method to estimate how accurately a predictive model will perform in practice. In a 5-fold cross-validation, the original sample is randomly split into 5 equal size sub-samples. Of the 5 sub-samples, a single sub-sample is left as the validation data for testing the model, and the remaining 4 sub-samples are used as training data. The cross-validation process is then repeated 5 times, with each of the 5 sub-samples used exactly once for the validation. The 5 results from the folds is averaged  to produce a single estimation of accuracy. The advantage of this method is that all observations are used for both training and validation, and each of them is used for validation only once.

In the benchmark scenario, we convert the daily open return into binary response by using zero as the cut-off point, as discussed in the data section. That is, if the open return is positive, $y_i=1$, otherwise $y_i=0$. The accuracy is defined as the rate of observations correctly classified using 0.5 as the cut point of the predicted probability $\pi$.

The prediction accuracies are shown in Table~\ref{tab:prediction}. The last two rows of Table~\ref{tab:prediction} show the result of the comparison experiments, in which the original data (functional data, compositional data, scalar data) are regarded as scalar predictors and are fed into the classification model. As can be seen, the prediction accuracies using the original data are indeed worse than those under our proposed framework. As a matter of fact, treating the functional data and compositional data as scalar predictors and directly imposing them together on statistic analysis can not be supported by statistical theory because they belong to different spaces.

Using the proposed framework, we have the accuracy of 0.65, 0.65, and 0.56 for the three phases, respectively. A substitution of the logistic regression is support vector machine (SVM), and the second row of Table~\ref{tab:prediction} shows the prediction accuracies that SVM provides. As can be seen, the SVM classifier (using a linear kernel) does not out-perform the logistic regression under the proposed framework. This result coincides with the previous study \citep{perlich2003tree}, in which the analysis of learning curves shows that logistic regression performs well for small data sets. Due to advantages of the good prediction accuracies and the convenience for interpretation of regression coefficients, the following discussion focus on the results that logistic regression yields.

\begin{table}[h]
\centering
\caption{Prediction accuracy.}\label{tab:prediction}
\begin{tabular}{rrlrrr}
  \hline
 & classification model & phase 1 & phase 2 & phase 3 \\
 \hline
\multirow{2}*{under the proposed framework} & logistic regression & \bf{0.65} & \bf{0.65} & \bf{0.56} \\
~ & SVM  & 0.65 & 0.59 & 0.50   \\
 \hline
 \multirow{2}*{using original data} & logistic regression & 0.54 & 0.44 & 0.46 \\
~ & SVM  & 0.50 & 0.57 & 0.47   \\
 \hline
\end{tabular}
\end{table}

The accuracies for the first two phases are higher than the previous study~\citep{zhou2017}, in which the logistic regression is also used based on only the emotion data and obtain an accuracy of 58.1\%, while the third phase failed to out-perform the previous study. The reason might be that in phase 3, the market kept being depressed and less information could be extracted from the variables we have. Nonetheless, the results have shown a good prediction capacity of the proposed framework based on multiple types of complex data.

\begin{figure}[h!]
	\begin{minipage}{0.5\linewidth}
		\centering
		\includegraphics[width =0.95\linewidth]{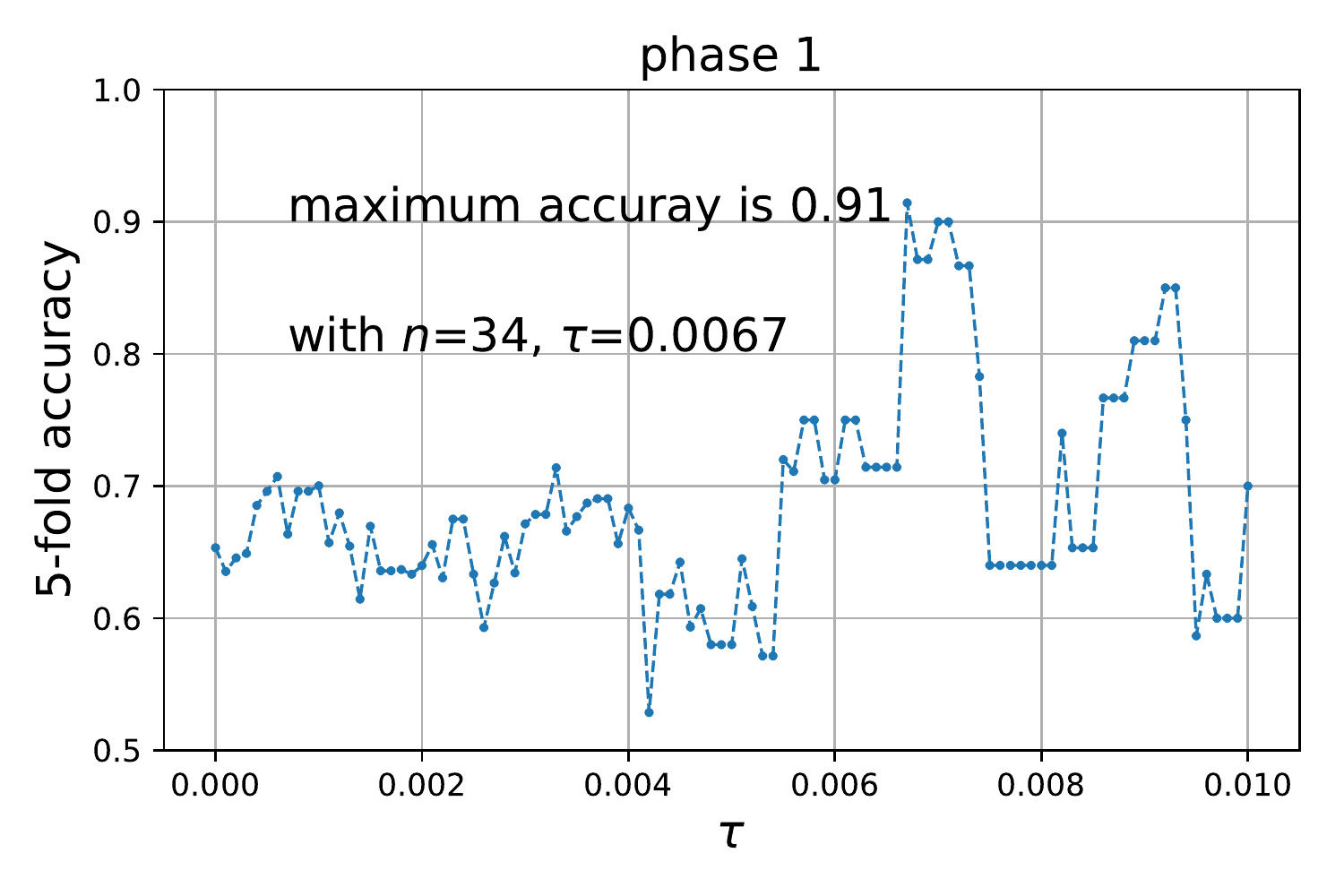}
	\end{minipage}\hfill
	\medskip
	\begin{minipage}{0.5\linewidth}
		\centering
		\includegraphics[width =0.95\linewidth]{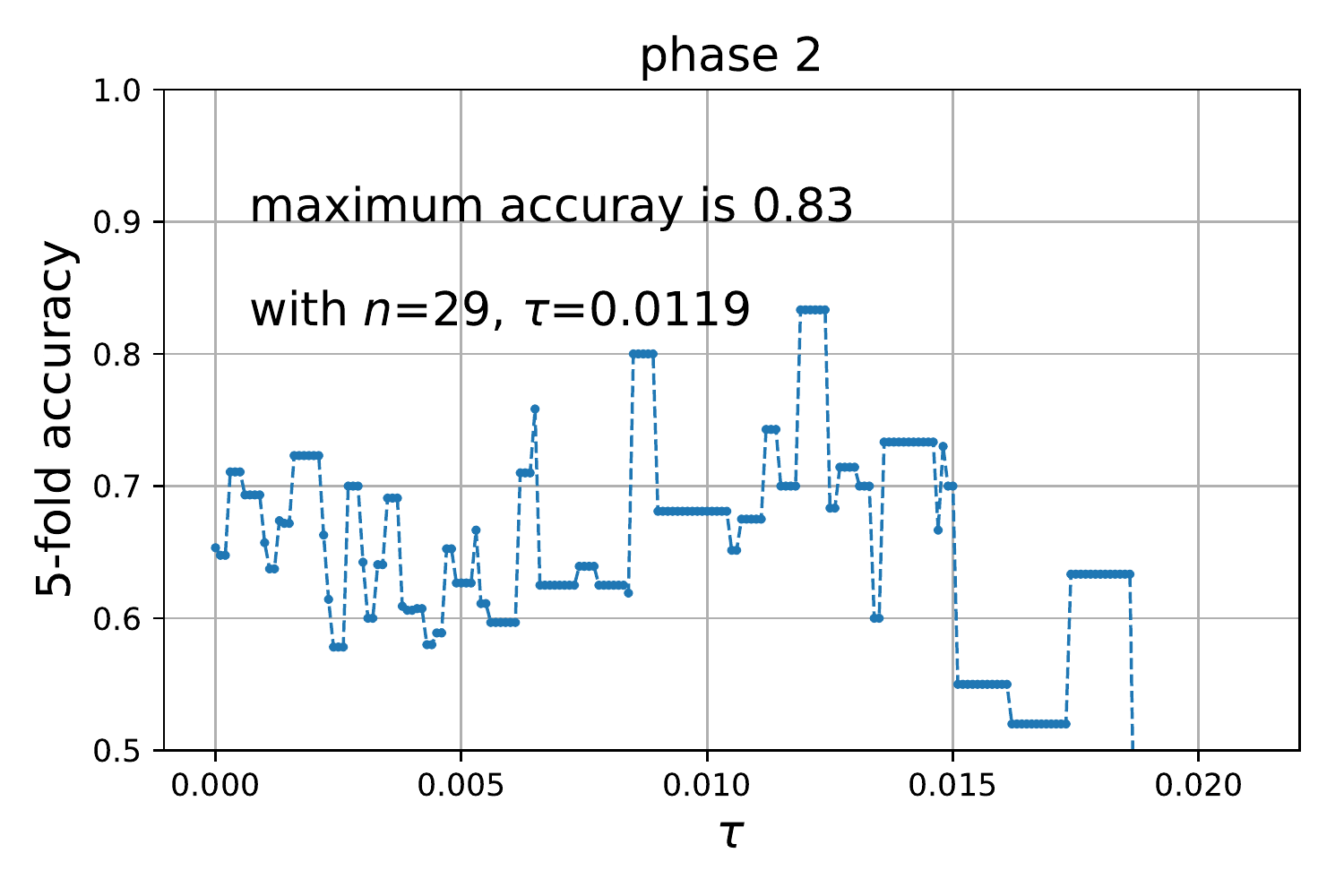}
	\end{minipage}
	\begin{minipage}{\linewidth}
		\centering
		\includegraphics[width =0.5\linewidth]{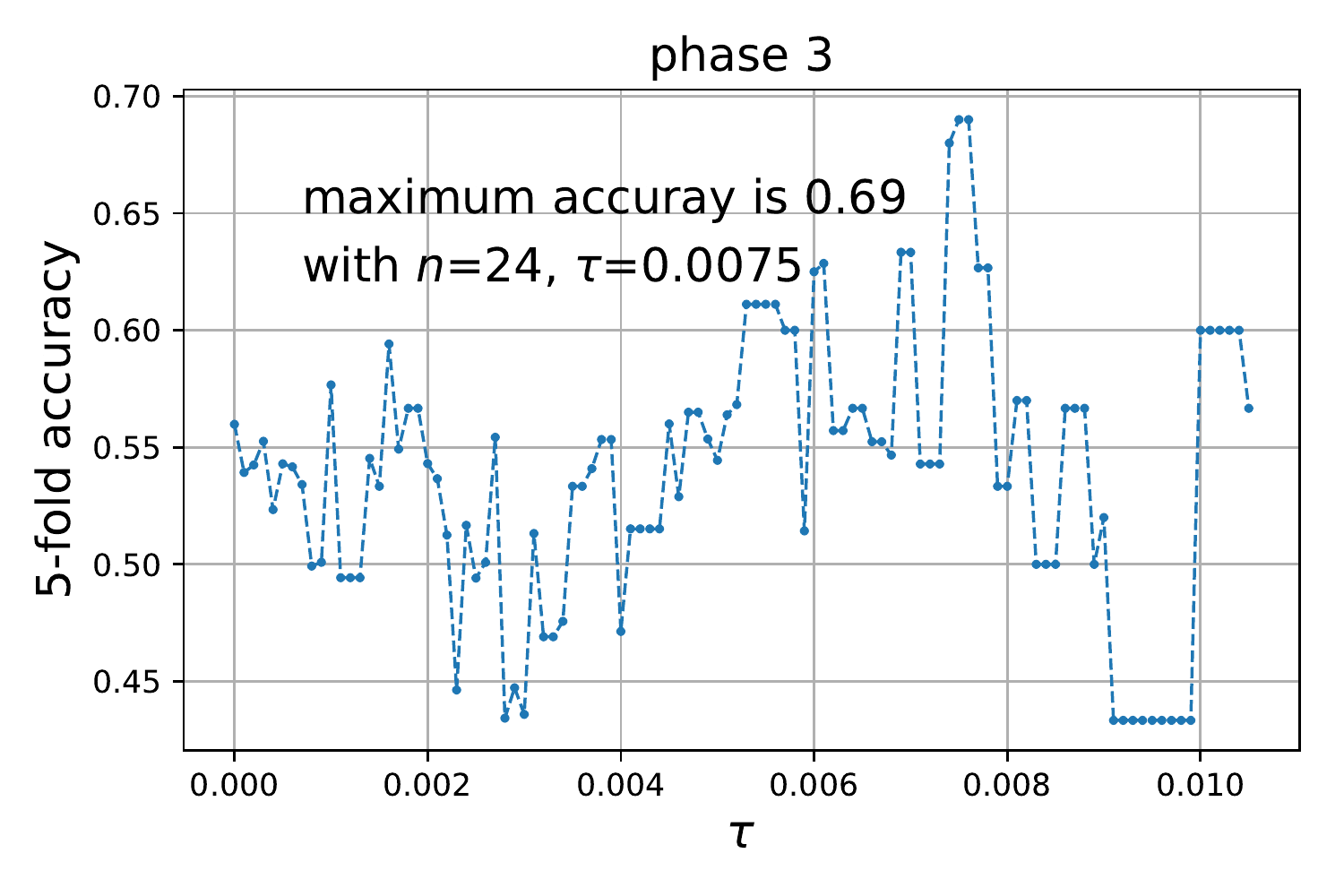}
	\end{minipage}
\caption{{\bf The relationship between the cut point $\tau$ and the 5-fold accuracy.} When $\tau=0$, the accuracy of is  0.65, 0.65, and 0.56 for the three phases as reported in Table \ref{tab:prediction}.}\label{fig:tao}
\end{figure}

At first glance of Table~\ref{tab:prediction}, the prediction accuracies of the proposed framework may not seem to be exciting. This is most likely due to the fact that, in the present modelling, we are using zero as the cut-off point of the daily open return, which is too sensitive to capture the significant difference between the ups and downs of the market. To weaken the sensitivity, we further present a threshold-based sampling approach. Define $\tau$ as the cut-off point of the daily open return, where $\tau \in [0,0.1]$. For any given $\tau$, we pick out the observation where the daily open return is higher than $\tau$ or lower than $-\tau$ and then perform the 5-fold cross validation of the prediction framework. Fig.~\ref{fig:tao} illustrates the relationship between $\tau$ and the accuracy. The maximum accuracies are also shown in the figures, with the corresponding sample sizes $n$ and thresholds $\tau$. As can be seen, the accuracy is sensitive to the threshold $\tau$. The trade-off here is that $\tau$ has successfully left the returns that of small absolute values out of consideration, while the number of observations is cut down, making the prediction accuracy unstable. Thus, the relationship between $\tau$ and accuracy is not monotonic. Nevertheless, Fig.~\ref{fig:tao} exhibits that the accuracy could be improved by selecting the significant ups or downs of the market in advance and sacrificing the number of observations, especially for phase 1 and phase 2, when the market experiencing big ups and downs. From this perspective, the framework is expected to reach better performance when the market vibrates sharply. Besides, Fig.~\ref{fig:tao} gives a comprehensive evaluation of the predictive power of the proposed framework.

\subsection{Coefficients interpretation}

For the three different sample periods, we apply the proposed method and obtain the estimated coefficients of compositional predictors (investors' emotion), functional predictors (series of the intraday 5-minutes returns), and scalar predictor (volume). The estimated coefficients of compositional predictors and scalar predictor are shown in Table~\ref{coefficent}. As can be seen, in the bullish market (phase 1), the most important emotion is ``sadness".
Though in a bullish market, the dominant emotion would be ``joy" (as shown by Fig.~\ref{fig:emotion}), the ``sadness" is the one that has the greatest influence on the opening return. In phase 2, the ``disgust" turns out to be the most influential one among the five kinds of emotion. This indicates that during the beginning of a bearish market, people's dislike for the market has the absolute priority to the market trending. In phase 3, where the market kept vibrating in depression, however, the ``anger" is of great importance to the opening return in the future, while other types of emotion plays their parts at the same time. It is worth noticing that ``fear" becomes important in phase 3 instead of phase 2, implying that it is not the fear at the beginning of the bearish market but the fear after the initial shock would strike the market trend.

\begin{table}[h]
\centering
\caption{Coefficients of emotions and volume.}\label{coefficent}
\begin{tabular}{rlrrr}
  \hline
 & phase 1 & phase 2 & phase 3 \\
  \hline
 anger & 0.01 & 0.00 & \bf{0.49} \\
 disgust & 0.04 & \bf{0.98} & 0.10 \\
 joy & 0.29 & 0.00 & 0.12 \\
 sadness & \bf{0.66} & 0.00 & 0.13 \\
 fear & 0.00 & 0.02 & 0.16 \\
 \hline
 volume & -0.84 & -0.79 & -0.24 \\
 \hline
\end{tabular}
\end{table}

Table~\ref{coefficent} also shows that the coefficient of volume is always negative. In fact, a high level of volume means the market participants hold diverge opinions on the market expectation, that is, some think it is time to sell while others believe that it is time to buy. Therefore, the negative coefficients imply that one unit higher of volume (or say, one unit of higher divergence of investors' opinions) happened yesterday will lead to some units lower of the probability for the stock index to perform a positive return at opening. And the absolute impact is decreasing from the bullish market to the bearish market.

\begin{figure}[h!]
\centering
\includegraphics[width=0.75\linewidth]{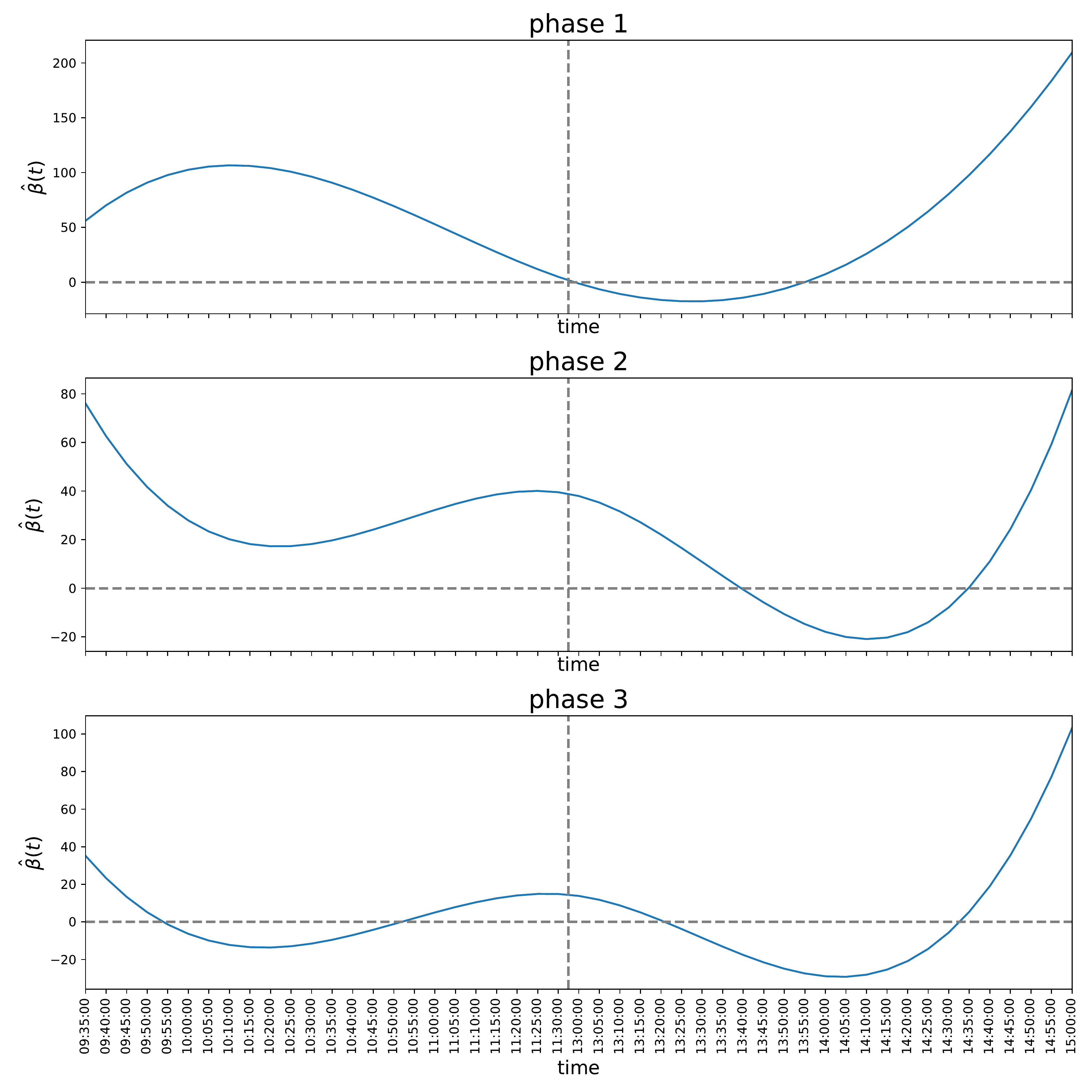}
\caption{{\bf Functional coefficients in the three phases.} The x-axis is the intraday time at 5 minutes frequency, from 9:35am to 11:30am and 13:00pm to 15:00pm. The curves illustrate the degree of impact of the intraday returns in the former day on the opening return today.}\label{fig:fun_coe}
\end{figure}

The estimated coefficients of functional predictors are shown in Fig.~\ref{fig:fun_coe}. The functional coefficients could be explained as the impact of the return at a specific time of yesterday on today's probability of opening return being positive. It illustrates how the intraday effect \citep{harris1986transaction,chang2008weather,foster1993variations} of stock prices influence the trending of the market during the boom-and-bust of Chinese stock market.

Specifically, Fig.~\ref{fig:fun_coe} illustrates the different characteristic of this impact in the three phases.
In the bullish market (phase 1), the impact increases to the top and goes down to zero before the closing in the morning. In the afternoon, the impact decreases to below zero at the beginning and goes up before closing.

During the initial market shock (phase 2), the impact of the intraday returns on the next day's probability of a positive open return decreases at the beginning of the day. Different from phase 1, the impact keeps being positive in the morning and the first half hour in the afternoon.
It goes down to negative in the afternoon and reverses back again to positive and flies to a high level before closing. The persistent positive impact throughout a day may be induced by the investors's persistent attention towards the market and the intraday returns in turn exert influence to the opening return in the next day.

Comparing to phase 1 and phase 2, phase 3 has the least of such kind of impact overall, as the functional coefficient is around zero in most of the time in the morning. This may source from the fact that the depressed state of the market could stir up little influence on the following day. It reaches its lowest around 14:00pm, similar with phase 2. This is just what the investors calls ``Magic 14:00" and ``Magic 14:30" that has terrified the market participants because the sharp falls always happened around 14:00pm to 14:30pm during the 2015 Chinese market crash. Our results further confirm that those special moments have a high level of negative impact on the following days.

The discussion in this section demonstrates the ability of the proposed framework to integrate multiple types of complex data and to explain how the variables work for the response.

\section{Conclusion}\label{sec:conclusion}
In this paper, we proposed a framework that aggregates three types of data forms, namely scalar variable, compositional variable, and functional variable, for predicting stock market. While the framework is model-independent, we mainly inspect the logistic regression in this study. In terms of isometric logratio transformation, functional principal component and logistic regression, we develop the estimation procedure of the framework in aggregating complex data. Numeric simulation experiments show that our proposed framework is effective.

In the empirical studies on Chinese stock market, the trading volume (scalar data), intraday return series (functional data), and investors' emotion (compositional data) from social media are used to predict whether the market is up or down at opening in the next day. By dividing the sample period into 3 phases, we find that the estimated coefficients of trading volume and intraday return series shed light on the different market status. Most surprisingly, we find that in the bullish market, the ``sadness" is more important to the future market trend than ``joy". In the initial collapse (phase 2), the ``disgust" plays a dominant role. When the market became depressed, the ``anger" and ``fear" begun to do their parts. Interestingly, our results show that it is not at the beginning of the bearish market but the subsequent period in which the investors' ``fear" comes to be indicative to market trend. Besides, our proposed method exhibits a competent prediction power, especially in the first two phases.

Though our proposed framework performs well in our study, it has inevitable limitations. For instance, the correlation among the time series is neglected as we treat the observations independently. Besides, the accuracy is not high enough for phase 3, where there's a depressed market. Future works could consider develop a panel data framework to solve the first limitation, and add other informative variables into the framework to solve the second.


\section{Acknowledgments}
This research was financially supported by National Natural Science Foundation of China (Grant No. 71420107025). ZJC thanks the National Key Research and Development Program of China (No.2016QY01W0205).

\section{Reference}

\end{document}